\documentclass[prb,twocolumn,floatfix,nobibnotes,footinbib,superscriptaddress,
showpacs]{revtex4}
\usepackage[latin1]{inputenc}
\usepackage{amsmath}
\usepackage{graphicx}

\newcommand{\vc}[1]{\ensuremath{\mathbf{#1}}}

\begin{document}

\title{Cluster Dual Fermion Approach to Nonlocal Correlations}

\author{H. Hafermann}
\affiliation{I. Institute of theoretical Physics, University of Hamburg, 20355
Hamburg, Germany}
\author{S. Brener}
\affiliation{I. Institute of theoretical Physics, University of Hamburg, 20355
Hamburg, Germany}
\author{A. N. Rubtsov}
\affiliation{Department of Physics, Moscow State University, 119992 Moscow,
Russia}
\author{M. I. Katsnelson}
\affiliation{Institute of Molecules and Materials, Radboud University,
6525 ED Nijmegen, The Netherlands}
\author{A. I. Lichtenstein}
\affiliation{I. Institute of theoretical Physics, University of Hamburg, 20355
Hamburg, Germany}

\date{\today}

\begin{abstract}
We formulate a general cluster Dual Fermion Approach to nonlocal
correlations in crystals. The scheme allows the treatment of long-range
correlations beyond cluster DMFT and nonlocal effects in realistic calculations
of multiorbital systems. We show that the the simplest approximation exactly
corresponds to free cluster DMFT. We further consider the relation between the
two-particle Green functions in real and dual variables. We apply this approach
by calculating the Green function of the Hubbard model in one dimension starting
from the two-site cluster DMFT solution. The result agrees well with the Green
function obtained from a DMRG calculation.
\end{abstract}


\maketitle

\section{INTRODUCTION}

One of the successful routes to the description of strongly correlated
systems is related to the Dynamical Mean Field Theory (DMFT)\cite{DMFT}. In this
scheme the system is mapped onto an effective local quantum impurity problem in
a self-consistently determined bath. The self-energy in the DMFT approach is
local in space but frequency dependent. However, there are many phenomena
for which non-local correlations are important and often correlations are
long-ranged. Among these are Luttinger-Liquid formation in low-dimensional
systems, non-Fermi-Liquid behavior due to van-Hove singularities in two
dimensions or d-wave pairing in high-$T_c$ superconductors. Obviously, DMFT is
not suitable for the description of such systems and there are cases in which
DMFT even fails qualitatively, as is for example the case for the
one-dimensional Hubbard model. For the description of these non-local
correlations it is desirable to combine short-range cluster many-body physics,
like formation of singlets, and long-range correlations. Recently the novel Dual
Fermion approach for the treatment of nonlocal correlations has been
developed\cite{Rub07}. Here we formulate a general cluster (or multi-orbital)
scheme for non-local correlations.

Frequently used approaches to account for non-local correlations
beyond DMFT comprise the cluster approaches or the so-called Dynamical Cluster
approximation (DCA) in k-space \cite{Het98}, real space periodic \cite{Li00}
and free cluster approaches \cite{Maz02}, as well as the Cellular-DMFT
\cite{Kot01}(CDMFT). In the latter approach, the single-site impurity of the
DMFT is replaced by a cluster of impurities. The CDMFT scheme however is
restricted to relatively small cluster sizes due to computational feasibility
and only short-range correlations can be treated within this approach\cite{DCA}.

Recently, steps have been taken to go beyond DMFT and to treat long-range
correlations. One of them is the Dynamical Vertex Approximation\cite{Tos07} and
similar approaches\cite{Kus06,Sle06}, where a diagrammatic expansion around DMFT
is made. A principally new scheme with a fully renormalized expansion called
Dual Fermion Approach has been proposed\cite{Rub07}, which is based on the
introduction of new variables in the path integral representation.
Similar schemes for the strong coupling expansion have been discussed in terms
of Hubbard operators \cite{Kot04}.

Here we present a cluster generalization of the Dual Fermion approach. This
allows the treatment of clusters or multiorbital atoms within the Dual Fermion
framework and can describe long-range correlations in realistic systems. The
paper is organized as follows: first, we present the formalism and then show
that for the case of non-interacting dual fermions one exactly obtains the
self-consistency equation of the CDMFT. In the next section we consider the
relationship between the two-particle Green functions in real and dual
variables. Then we describe the calculation procedure. We finally apply our
approach to the Hubbard model in one dimension starting from a two-site CDMFT
solution.

\section{Formalism}

Our goal is to find an (approximate) solution to the cluster lattice problem
described by the imaginary time action which in cluster or multiband notation
reads
\begin{eqnarray}
S[c^\ast,c]\!&=&\!\!
-\!\!\!\!\!\sum_{\omega\vc{k}\sigma
mm^\prime}\!\!\! c^\ast_{\omega\vc{k}\sigma m}\left((i\omega+\mu)\vc{1}
-H_{\sigma}(\vc{k})\right)_{mm^\prime}c_{\omega\vc{k}\sigma m^\prime}\nonumber\\
&+& \sum_{i} H_{\text{int}}[c_i^\ast,c_i]\ .
\end{eqnarray}
Here $H_{\sigma}(\vc{k})$ is the one-electron part of the
Hamiltonian, $\omega =(2n+1)\pi /\beta, n=0,\pm 1,...$ are the
Matsubara frequencies, $\beta $ and $\mu $ are the inverse
temperature and chemical potential, respectively, $\sigma
=\uparrow ,\downarrow $ labels the spin projection and $c^{\ast
},c$ are Grassmannian variables. The indices $i$ \ label the
translations of the super-cell and the $\vc{k}$-vectors span the
reciprocal lattice in the reduced super-cell Brillouin zone. Here
we take into account a general type of interaction,
$H_{\text{int}}$. It is important to note that it can be
\emph{any} type of interaction inside the multiorbital atom or
cluster. The only requirement is that $H_{\text{int}}$ is local.
For example, the general Coulomb interaction has the form
\begin{equation}
H_{\text{int}}[c_i^\ast,c_i] = \frac{1}{4}\sum_{i}
\int\limits_0^\beta d\tau\, U_{1234} c^\ast_1 c^\ast_2 c_4 c_3\ ,
\end{equation}
where $U$ is the general symmetrized Coulomb vertex and e.g.
$1\equiv\{\omega_1 m_1\sigma_1\}$ comprehends frequency-, orbital- and spin
degrees of freedom and summation over these states is implied. In order to
capture the
local physics we introduce a cluster impurity problem just in the spirit of
CDMFT in the form
\begin{eqnarray}
S_{\text{imp}}[c^\ast,c]\!&=&\!\!
-\!\sum_{\omega\sigma}
c^\ast_{\omega\sigma
m}\left((i\omega+\mu)\vc{1}-\Delta_{\omega\sigma}\right)_{mm^\prime} c_{
\omega\sigma m^\prime}\nonumber\\
&+& H_{\text{int}}[c^\ast,c]\ ,
\end{eqnarray}
where $\Delta$ is an as yet unspecified hybridization function describing the
interaction of the impurity with an electronic bath.
We suppose that all properties of the impurity problem are in principle known,
i.e. the single-particle Green function $g_{\sigma\omega}$ and
the irreducible vertices $\gamma ^{(4)},\gamma ^{6},$ etc. are known. Our
goal is to express the Green function $G_{\omega k}$ and vertices $\gamma$ of
the original lattice problem via these quantities.

Since $\Delta$ is local, one may formally rewrite the original lattice problem
in the following form:
\begin{eqnarray}
S[c^\ast,c]&=&\sum_i S_{\text{imp}}[c^\ast_{\omega
i\sigma},c_{\omega i\sigma}]\nonumber\\
&-& \sum_{\omega\vc{k}\sigma mm^\prime} c^\ast_{\omega\vc{k}\sigma m}
\left(\Delta_{\omega\sigma} -
H_{\sigma}(\vc{k})\right)_{mm^\prime}\vc{c}_{\omega\vc{k}\sigma
m^\prime}\ .\nonumber\\
\label{eq::action_rew}
\end{eqnarray}
We introduce spinors $\vc{c}_{\omega\vc{k}\sigma}=(\ldots
,c_{\omega\vc{k}\sigma m},\ldots)$, $\vc{c}^\ast_{\omega\vc{k}\sigma}=(\ldots
,c^\ast_{\omega\vc{k}\sigma m},\ldots)$. Omitting indices, in matrix-vector
notation, the Gaussian identity that facilitates the transformation to the dual
variables is
\begin{eqnarray}
&&\int \exp\left(-\vc{f}^*\hat{A}\vc{f} -
\vc{f}^*\hat{B}\vc{c}
- \vc{c}^*\hat{B}\vc{f}  \right) \mathcal{D}[\vc{f},\vc{f}^*] =\nonumber\\
&&\det(\hat{A})\ \exp\left(\vc{c}^*\hat{B}\hat{A}^{-1}\hat{B}\vc{c}\right)\ ,
\end{eqnarray}
which is valid for arbitrary complex matrices $\hat{A}$ and $\hat{B}$. In order
to decouple the non-local term in Eqn. \ref{eq::action_rew}, we choose
\begin{eqnarray}
A&=&
g_{\omega\sigma}^{-1}\left(\Delta_{\omega\sigma}-H_\sigma(\vc{k})\right)^{-1}g_{
\omega\sigma}^{-1}\ ,\nonumber\\
B&=&g_{\omega\sigma}^{-1}\ ,
\end{eqnarray}
where $g_{\omega\sigma}$ is the Green function matrix in the
orbital space $(m,m^\prime)$ of the local impurity problem. Using
this identity, the lattice action can be rewritten in the form
\begin{eqnarray}
&&S[\vc{c}^*,\vc{c},\vc{f}^*,\vc{f}] = \sum_i S_{\text{site},i} +\nonumber\\
&&\sum_{\omega\vc{k}\sigma}\left[\vc{f}^*_{\omega\vc{k}\sigma}\
g_{\omega\sigma}^{-1}\ \left(\Delta_{\omega\sigma} -
H_{\sigma}(\vc{k})\right)^{-1}\
g_{\omega\sigma}^{-1}\ \vc{f}_{\omega\vc{k}\sigma}\right]\ ,
\end{eqnarray}
where
\begin{equation}
\sum_i S_{\text{site},i}= \sum_i S_{\text{imp}}[\vc{c}^*_i,\vc{c}_i] +
\vc{f}^*_{\omega i\sigma}\ g_{\omega\sigma}^{-1}\vc{c}_{\omega i\sigma} +
\vc{c}^*_{\omega i\sigma}\ g_{\omega\sigma}^{-1}\vc{f}_{\omega i\sigma}\ .
\label{eq::Ssite}
\end{equation}
Here the summation in the last term over states labeled by
$\vc{k}$ has been replaced by the equivalent summation over all
sites. The Gaussian identity can further be used to establish an
exact relation between the lattice Green function and the dual
Green function. To this end, the partition function of the lattice
is written in the two equivalent forms
\begin{eqnarray}
Z&=&\int\exp\left(-S[\vc{c}^\ast,\vc{c}]\right) \mathcal{D}[\vc{c},\vc{c}^\ast]
=\nonumber\\ && Z_f \int\int
\exp\left(-S[\vc{c}^\ast,\vc{c},\vc{f}^\ast,\vc{f}]\right)
\mathcal{D}[\vc{f},\vc{f}^\ast]\mathcal{D}[\vc{c},\vc{c}^\ast]\ ,\nonumber\\
\label{eq::partitionfunction}
\end{eqnarray}
where
\begin{equation}
Z_f = \prod\limits_{\omega\vc{k}\sigma}
\det\left[g_{\omega\sigma}\left(\Delta_{\omega\sigma}-H_\sigma(\vc{k}
)\right)g_{\omega\sigma}\right]\ .
\end{equation}
By taking the functional derivative of the partition function,
Eqn. \ref{eq::partitionfunction}, with respect to the Hamiltonian, i.e.
\begin{equation}
G_{\omega\vc{k}\sigma}^{mm^\prime}= \frac1Z\frac{\delta Z}{\delta
H_{\sigma} (\vc{k})_{m^\prime m}}\ ,
\end{equation}
one can obtain the following exact relationship between the dual and lattice
Green functions:
\begin{eqnarray}
G_{\omega\vc{k}\sigma} &=& \left(g_{\omega\sigma} \left(\Delta_{\omega\sigma} -
H_{\sigma}(\vc{k})\right)\right)^{-1}
G^{\text{d}}_{\omega\vc{k}\sigma} \times \nonumber \\
&\times& \left(\left(\Delta_{\omega\sigma} -
H_{\sigma}(\vc{k})\right) g_{\omega\sigma}\right)^{-1}
+ \left(\Delta_{\omega\sigma} - H_{\sigma}(\vc{k})\right)^{-1}\ ,\nonumber \\
\label{eq::Glat}
\end{eqnarray}
where the lattice Green function is defined via the imaginary time
path integral as
\begin{equation}
G_{12} = -\frac{1}{Z}\int c_{1}c_{2}^\ast
\exp\left(-S[\vc{c}^\ast,\vc{c}]\right)
\mathcal{D}[\vc{c},\vc{c}^\ast]
\end{equation}
and similarly for the local Green function $g$ and dual Green function $G^d$
with $Z$ and $S$ replaced by the corresponding expressions.


We now wish to derive an action depending on the dual variables only. This can
be achieved by integrating out the original variables $\vc{c}$,$\vc{c}^\ast$.
The crucial point is that this can be done for each site separately:
\begin{eqnarray}
&&\int\exp\left(-S_{\text{site}}[\vc{c}_i^\ast,\vc{c}_i,\vc{f}_i^\ast,\vc{f}_i]
\right)\mathcal{D}[\vc{c}_i,\vc{c}_i^\ast] =\nonumber \\
&& Z_{\text{imp}}e^{-\left(\sum_{\omega\sigma} \vc{f}^\ast_{\omega i\sigma}\
g^{-1}_{\omega\sigma}\ \vc{f}_{\omega i\sigma} +
V_i[\vc{f}_i^\ast,\vc{f}_i]\right)}\ .
\end{eqnarray}
This equation can be seen as the defining equation for the dual potential
$V[\vc{f}^\ast,\vc{f}]$. Since $S_{\text{site}}$ contains the impurity action,
expanding the remaining part of the exponential and integrating out the original
variables
corresponds to averaging over the impurity degrees of freedom. Equating
the resulting expressions by order, one finds that the dual potential in lowest
order approximation is given by
\begin{equation}
V[\vc{f}^\ast,\vc{f}] = \frac{1}{4} \sum\limits_i \gamma^{(4)}_{1234}
\vc{f}_{i1}^\ast
\vc{f}_{i2}^\ast \vc{f}_{i4} \vc{f}_{i3} + \ldots
\end{equation}
where
\begin{eqnarray}
\gamma^{(4)}_{1234} &=& g_{11^\prime}^{-1}g_{22^\prime}^{-1} \left[
\chi^{\text{imp}}_{1^\prime 2^\prime 3^\prime 4^\prime} -
\chi^{(0)\text{imp}}_{1^\prime 2^\prime 3^\prime 4^\prime} \right]g_{3^\prime
3}^{-1}g_{4^\prime 4}^{-1}\ , \nonumber \\
&&\chi^{(0)\text{imp}}_{1234} = g_{14}g_{23}
- g_{13} g_{24}
\end{eqnarray}

is the fully antisymmetric irreducible vertex and the local
two-particle Green function of the impurity model is defined as
\begin{equation}
\chi^{\text{imp}}_{1234}=\frac{1}{Z_{\text{imp}}}\int
c_{1}c_{2} c_{3}^\ast c_{4}^\ast
\exp\left(-S_{\text{imp}}[\vc{c}^\ast,\vc{c}]\right)
\mathcal{D}[\vc{c},\vc{c}^\ast]\ .
\end{equation}

\begin{figure}[t]
\begin{center}
\begin{tabular}{ccc}
\includegraphics[scale=0.12,angle=0]{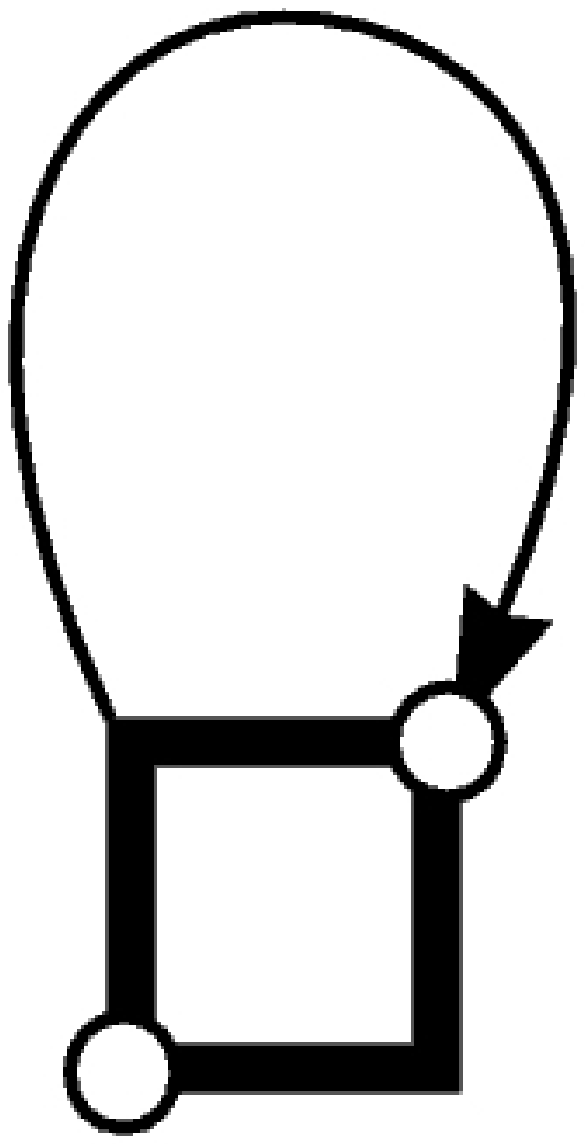} & \hspace{4em}
\includegraphics[scale=0.12,angle=0]{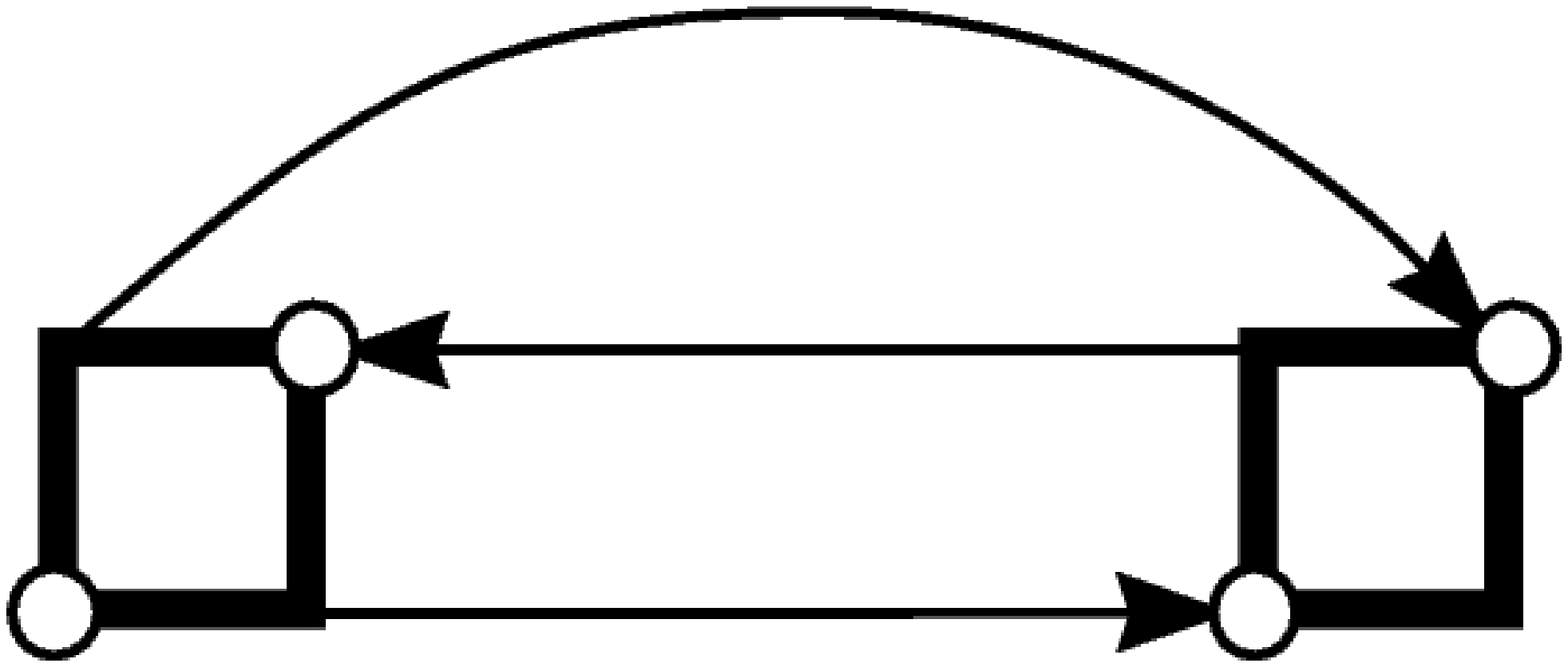}
\end{tabular}
\end{center}
\caption{The first two lowest order diagrams for the dual
self energy $\Sigma^d$.}\label{fig::diagrams}
\end{figure}

The dual action now depends on dual variables only and can be written as
\begin{equation}
S_{\text{d}}[\vc{f}^\ast,\vc{f}] = -\sum_{\omega\vc{k}\sigma}
\vc{f}^\ast_{\omega\vc{k}\sigma}
\left(G^{\text{d}(0)}_{\omega\vc{k}\sigma}\right)^{-1}
\vc{f}_{\omega\vc{k}\sigma} + \sum_i V[\vc{f}_i^\ast,\vc{f}_i]\ .
\label{eq::dual_action}
\end{equation}
The bare dual Green function is given by
\begin{equation}
G^{\text{d}(0)}_{\omega\vc{k}\sigma} =
-g_{\omega\sigma}\left[\left(\Delta_{\omega\sigma} -
H_{\sigma}(\vc{k})\right)^{-1} + g_{\omega\sigma} \right]^{-1}
g_{\omega\sigma}\ .
\label{eq::Gdbare}
\end{equation}
For the nonlocal part of the self energy defined by
$\Sigma_{\text{nonloc}} = \Sigma - \Sigma_{\text{loc}}$ we find a
simple matrix relation to the dual self-energy:
\begin{equation}
 \Sigma_{\text{nonloc}}^{-1} = \Sigma_{\text{d}}^{-1} + g\ .
\end{equation}
The local part of the self energy is obtained by the solution of the
impurity problem. In order to obtain the nonlocal contribution, we thus need to
calculate the dual self-energy. This is achieved by performing a regular
diagrammatic series expansion of the dual action, Eqn. \ref{eq::dual_action} and
considering the lowest order diagrams for $\Sigma_{\text{d}}$, constructed
from the irreducible vertices and the dual Green function as lines. The diagrams
considered here are shown in Fig. \ref{fig::diagrams}. The lowest order diagram
is local while the next diagram already gives a nonlocal contribution to the
self energy.

So far we have not established a conditon for $\Delta$. We require that the
first diagram in the expansion of the dual self-energy should be equal to zero
at all frequencies. Since $\gamma^{(4)}$ is local, we can use the following
condition:
\begin{equation}
\sum_{\vc{k}}G_{\vc{k}\omega }^{d}=0\ .
\end{equation}
In the simplest approximation, which corresponds to non-interacting dual
fermions, the full dual Green function is replaced by the corresponding bare
Green function and the above condition can be reduced to
\begin{equation}
\sum_{\vc{k}} \left[\left(\Delta_{\omega\sigma} -
H_{\sigma}(\vc{k})\right)^{-1} + g_{\omega\sigma} \right]^{-1}=0
\end{equation}
which is equivalent to the self-consistency condition for the
hybridization function in free-cluster CDMFT
\cite{Li00,Kot01,Maz02}. The self-consistency condition for $\Delta$ which
satisfies
this condition is
\begin{equation}
\Delta_{\text{new}} = \Delta_{\text{old}} + g^{-1} G^\text{d}_{\text{loc}}
G^{-1}_{\text{loc}}\ .
\label{eq::selfc}
\end{equation}
If we restrict the matrix $\Delta_{\omega\sigma}$ to be equivalent for all
cluster sites with periodic boundary conditions then this approximation leads to
the DCA scheme \cite{DCA}. 

\section{Two-particle excitations}

In order to find the exact relation between the four-point
correlation function in dual and conventional variables we have to
calculate the second derivative of $Z$ with respect to
$H_{\sigma}(\vc{k})_{\mu\lambda}$, $H_{\sigma}(\vc{k})_{\rho\nu}$
using the two equivalent representations of the partition
function, Eqn. \ref{eq::partitionfunction}. Differentiating the
partition function in its original form gives us
\begin{equation}
\frac1Z\frac{\delta^2 Z}{\delta H_{\mu\lambda}\delta H_{\rho\nu}}
=\langle \mathrm
Tc_{\lambda}c_{\mu}^{\dagger}c_{\nu}c_{\rho}^{\dagger}\rangle\equiv\chi_{
\lambda\mu\nu\rho}\ ,
\end{equation}
while differentiating the second expression for $Z$ using the
above relation for Green's function, Eqn. \ref{eq::Glat}, yields
for $\chi_{\lambda\mu\nu\rho}$ after some straightforward algebra:
\begin{multline}
\label{chi-chid} \frac1Z\frac{\delta^2 Z}{\delta H_{ \nu\mu}\delta
H_{\rho\lambda}}
=\left[(\Delta-H)^{-1}\otimes(\Delta-H)^{-1}\right]_{\lambda\mu\nu\rho}+\\
\left[(\Delta-H)^{-1}\otimes[(\Delta-H)^{-1}g^{-1}G_dg^{-1}(\Delta-H)^{-1}]
\right]_{\lambda\mu\nu\rho}+\\
\left[[(\Delta-H)^{-1}g^{-1}G_dg^{-1}(\Delta-H)^{-1}]\otimes(\Delta-H)^{-1}
\right]_{\lambda\mu\nu\rho}+\\
\left[(\Delta-H)^{-1}g^{-1}\right]_{\lambda
l}\left[(\Delta-H)^{-1}g^{-1}\right]_{\nu
n}\chi_{lmnr}^d \times \\
\times
\left[g^{-1}(\Delta-H)^{-1}\right]_{m\mu}\left[g^{-1}(\Delta-H)^{
-1}\right]_{r\rho}.
\end{multline}
Here $\chi_{lmnr}^d\equiv\langle\mathrm
Tf_{l}f_{m}^{\dagger}f_{n}f_{r}^{\dagger}\rangle$ is the dual
four-point correlation function and
$$
(A\otimes B)_{\lambda\mu\nu\rho}\equiv
A_{\lambda\mu}B_{\nu\rho}-A_{\lambda\rho}B_{\nu\mu}
$$
is the antisymmetrized direct product of two matrices.

As one can see from (\ref{chi-chid}), the two-particle excitations
for dual fermions coincide with those for real fermions. Thus to
get the information about the instabilities in the system under
consideration it is sufficient to sum up the ladder for the
two-particle dual fermion Green function in the ladder
approximation:
\begin{equation}
\chi_{lmnr}^{d}=\chi_{lmnr}^{d0}+\chi_{l\mu
n\rho}^{d0}\gamma_{\mu\rho\lambda\nu}\chi_{\lambda m\nu r}^{d}
\end{equation}
or
\begin{equation}
\chi_{lmrn}^{d}=\chi_{lmrn}^{d0}+\chi_{l\mu \rho
n}^{d0}\gamma_{\mu\nu\rho\lambda}\chi_{\lambda mr\nu}^{d}
\end{equation}
where the first equation is written for the particle-particle
channel and the second for the particle-hole one, and
$\chi^{d0}=G_d\otimes G_d$ is the bare two-particle dual Green
function.

\section{Calculation procedure}

\begin{figure}[t]
\begin{center}
\includegraphics[scale=0.25,angle=0]{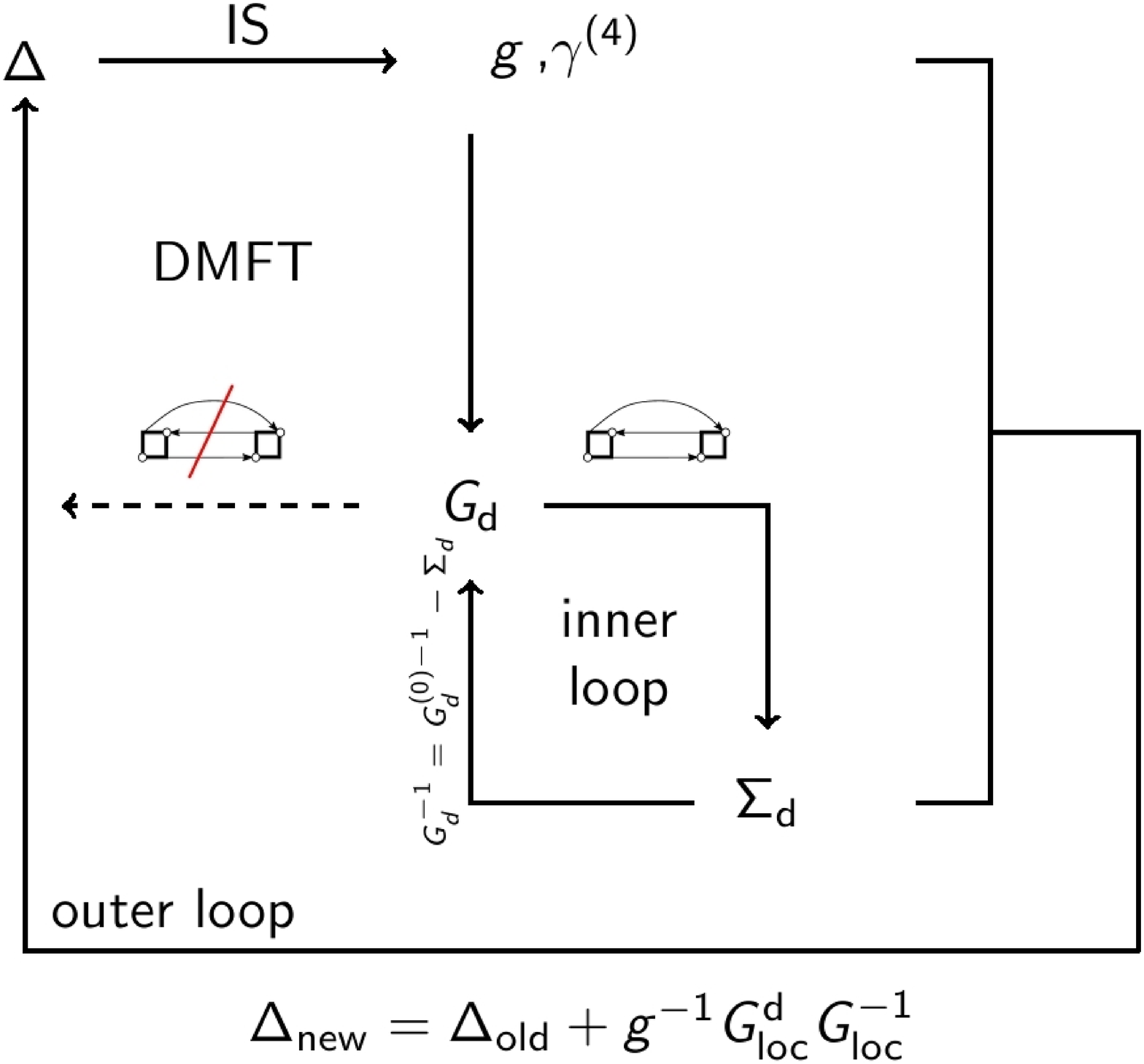}
\end{center}
\caption{Calculation procedure}\label{fig::procedure}
\end{figure}

The calculation procedure is as follows:
Starting from a starting guess for $\Delta$, e.g. $\Delta=0$, we solve the
impurity problem and obtain a new local Green function $g$. From this we
calculate the bare dual $\vc{k}$-dependent Green function via Eqn.
\ref{eq::Gdbare}. In order to reach the self-consistent free-cluster DMFT
solution we calculate the local part of the lattice Green function via Eqn.
\ref{eq::Glat} and insert it into Eqn. \ref{eq::selfc} to obtain a new $\Delta$
with which we again solve the impurity problem. This loop, which is closed by
the dashed arrow in Fig. \ref{fig::procedure}, is repeated until the
self-consistent CDMFT solution is reached. With the thus obtained $\Delta$ we
then calculate the irreducible vertex $\gamma^{(4)}$. Now we do not follow the
path indicated by the dashed line but calculate an approximation to the dual
self-energy by summing up the first diagram(s) in the perturbation series
expansion. From this and the bare dual Green function an approximation for the
dual Green function $G^{\text{d}}_{\omega\vc{k}\sigma}$ is obtained which is
subsequently used in the diagrams. This inner loop is executed until
self-consistency. The dual Green function and the lattice Green function
are then again used to obtain a new hybridization function, which serves as
input for the calculation of a new local Green function and renormalized vertex
in the impurity solver step. This outer loop is also executed until
self-consistency. Self-consistency for both loops is usually reached after a few
iterations (depending on the system). The computational cost for the
calculations aside from DMFT is less than for the DMFT itself, whereby the
computation of the vertex is the computationally most expensive part. We use the
continuous-time quantum Monte Carlo impurity solver\cite{Rub05} for the solution
of the impurity problem and for the calculation of the irreducible vertex.

\section{Application to 1D Hubbard model}

We consider the one-dimensional Hubbard model described by the
following Hamiltonian:
\begin{equation}
H_0+H_{\text{int}}=t \sum\limits_i c_{i+1}^\dagger c_i + U
\sum\limits_i n_{i\uparrow}n_{i\downarrow}\ .
\end{equation}

\begin{figure}[h]
\begin{center}
\vspace{1em}
\includegraphics[scale=0.45,angle=0]{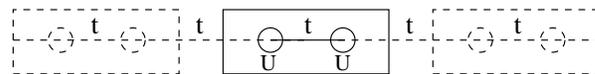}
\end{center}
\caption{Schematic representation of the 1D chain as a chain
of two-site clusters.}\label{fig::chain}
\end{figure}

For the half-filled case the main physics is related with the formation
of a local singlet pair. Therefore we start from a two-site cluster DMFT
solution and then include the long-ranged non-local effects via the Cluster Dual
Fermion approach. When this one-dimensional system is treated as a chain of
two-site clusters as depicted in Fig. \ref{fig::chain}, the tight-binding
Hamiltonian
for this model is readily shown to be
\begin{equation}
H_0(k) = \left(\begin{array}{cc}
0 & t(1+e^{-ik a})\\
t(1+e^{ika }) & 0
\end{array}\right)\ .
\end{equation}

\begin{figure}[t]
\begin{center}
\includegraphics[scale=0.85,angle=0]{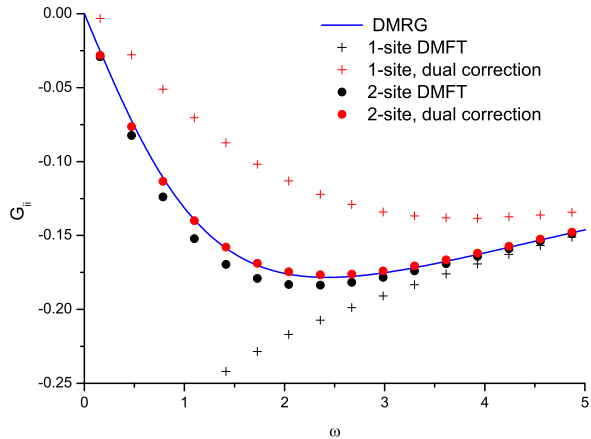}
\end{center}
\caption{(color online) Local Matsubara Green function on the
Matsubara axis obtained from DMRG for $T=0$ in comparison with the results
obtained from DMFT and from fully self-consistent dual fermion calculations.
For the 2-site free cluster DMFT the renormalization of the vertex has a small
effect since the CDMFT is already a good starting point. For the single site
calculation the renormalization is essential since DMFT even gives a
qualitatively wrong answer, while the dual fermion result
correctly predicts the system to be an insulator.}\label{fig::results}
\end{figure}

Due to the absence of a Mott transition in one dimension, the
system is an insulator for any finite value of the on-site
repulsion $U$. It is known that single-site DMFT gives a
qualitatively wrong answer, i.e. predicts the system to be
metallic even for $U$ as large as $U/t=6$. Here we compare our
results with the one obtained by a Density Matrix Renormalization
Group (DMRG) calculation \cite{DMRG}, since DMRG is known to
reproduce the spectral properties of 1D systems quite well. The
free cluster DMFT solution obtained in our calculation is
qualitatively correct and reproduces the DMRG solution quite well,
since short range fluctuations between nearest neighbours are
predominant. For our calculations we use the parameters $U/t=6$
and $\beta=20$. The DMRG solution corresponds to $T=0$. For the
calculation we considered only the first two lowest order diagrams
in the series expansion shown in Fig. \ref{fig::diagrams}. The
result of our calculation is depicted in Fig. \ref{fig::results},
where we show the imaginary part of the on-site Green function as
a function of Matsubara frequencies on one of the cluster sites.
In this calculation the renormalization of the vertex has a small effect, since
the CDMFT result is already close to the exact solution. However, the free
cluster DMFT result is considerably improved and the result agrees very well
with the one obtained in the DMRG calculation. Almost the same result was
obtained without changing the vertex.
We further plot the result for a single-site calculation. Here single-site DMFT
wrongly predicts the system to be metallic. Hence the DMFT solution strongly
differs from the exact solution and the renormalization of the vertex becomes
important. Nevertheless, the dual fermion solution converges to a metallic
solution within a few iterations.

\section{Conclusions}
We have generalized the recently proposed Dual Fermion Approach to
the multiorbital case, facilitating the treatment of multiorbital systems within
this framework.
We further established the relation between the four-point correlation functions
in real and dual variables and found that the two-particle excitations for real
fermions and dual fermions coincide. The approach was applied to the
one-dimensional Hubbard model starting from the free two-site cluster DMFT
solution. Although the CDMFT solution already quite well reproduces the DMRG
solution, the cluster dual fermion solution considerably improved this result.
The cluster formulation allows to combine this approach with realistic density
functional calculations and thus opens a new way to describe long-range
correlations in real systems.

\begin{acknowledgments}
The work was supported by NWO project 047.016.005 and FOM (The
Netherlands), DFG Grant No. SFB 668-A3 (Germany), the Leading
scientific schools program and the ``Dynasty'' foundation (Russia).
\end{acknowledgments}

\end{document}